# Electric Field Effect in Diluted Magnetic Insulator Anatase Co:TiO$_2$

T. Zhao[1,*], S. R. Shinde[2], S. B. Ogale[1,2,#], H. Zheng[1], T. Venkatesan[2]

[1]Department of Materials Science and Engineering, University of Maryland, College Park, MD 20742.

[2]Center for Superconductivity Research, Department of Physics, University of Maryland, College Park, MD 20742.

R. Ramesh[+]

Department of Physics and Department of Materials Science and Engineering, University of California, Berkeley, CA 94720.

S. Das Sarma

Condensed Matter Theory Center, Department of Physics, University of Maryland, College Park, MD 20742.

Abstract

An external electric field induced reversible modulation of room temperature magnetic moment is achieved in an epitaxial and insulating thin film of dilutely cobalt-doped anatase $TiO_2$. This first demonstration of electric field effect in any oxide based diluted ferromagnet is realized in a high quality epitaxial heterostructure of $PbZr_{0.2}Ti_{0.8}O_3$/Co:$TiO_2$/$SrRuO_3$ grown on (001) $LaAlO_3$. The observed effect, which is about 15% in strength in a given heterostructure, can be modulated over several cycles. Possible mechanisms for electric field induced modulation of insulating ferromagnetism are discussed.

The rapidly developing field of spin electronics embodies the notion of synergetic and multifunctional use of charge and spin dynamics of electrons, aiming to transcend the traditional dichotomy of semiconductor electronics and magnetic storage technology.[1] Such new functionalities require new materials encompassing both semiconducting and magnetic properties as vehicles to implement novel device concepts. An important approach to the development of such new materials is to examine the possibility of magnetizing a functional non-magnetic host by dilute magnetic-impurity doping which allows the host to retain its other desirable properties.[2] The corresponding materials are classified as diluted magnetic semiconductors (DMS). Considerable success has been achieved in this direction in the domain of III-V and group IV semiconductors[3-5], with some recent success reported in oxide-based systems[6-11]. Unfortunately in the case of many systems, researchers have not yet been able to completely rule out the possibilities of extrinsic effects such as dopant clustering, impurity magnetic phases etc.[12,13], and it has not always been clear that the reported DMS ferromagnetism is indeed induced by the carriers.

It is generally suggested that any new material must satisfy three specific tests to qualify as an intrinsic (carrier-induced, uniformly dispersed dopants) diluted ferromagnetic system: Observation of a) Anomalous Hall effect (AHE)[14,15], b) optical magnetic circular dichroism (O-MCD)[16,17], and c) electric field induced modulation of magnetization[18,19]. We have recently shown that observation of AHE, although desirable, is not really a strict test and can be realized even in a sample with non-percolating magnetic clusters.[20,21] The O-MCD test has been demonstrated in different systems including some oxide-based materials.[16,17,22] The electric field modulation, which in some sense is the most stringent test of intrinsic ferromagnetism, has, however, been demonstrated only in III-V DMS systems[18,19], but not yet in any oxide-DMS materials. In this paper we report the first successful implementation of an external electric field modulation of ferromagnetism in DMS anatase Co:$TiO_2$, grown under conditions suggested to yield uniform matrix incorporation of cobalt. We should however like to point out that the corresponding material is insulating in nature.

The sample structure employed in this study is shown in the inset in Fig.1. The basic methods of growing and testing such heterostructure devices are similar to those used in our previous studies on manganites[23-26]. A bottom electrode layer $SrRuO_3$ (SRO), a DMS anatase Co:$TiO_2$ layer (7 at% Co) and a ferroelectric $PbZr_{0.2}Ti_{0.8}O_3$ (PZT) layer were grown on $LaAlO_3$ (LAO) substrate by pulsed laser deposition (PLD). The temperature and oxygen pressure during growth were 650°C and $10^{-1}$ Torr for SRO and PZT, and 875°C and $3 \times 10^{-5}$ Torr for Co:$TiO_2$, respectively. A top Pt layer was deposited by PLD as well. A 200μm×200μm electrode size and a 50% coverage of the surface were achieved by standard photolithography and lift-off process.

The choice of the growth conditions for Co:$TiO_2$ was guided by the need to achieve a potentially intrinsic DMS system without major concerns about clustering of the dopant that are widely debated in the literature. Shinde et al.[7] showed that high temperature annealing of anatase Co:$TiO_2$ grown at 700 $^0$C, dissolves the cobalt clusters formed in the as-grown films. Taking a cue from this observation, films were directly grown at a higher substrate temperature (875°C) and examined by various techniques to study lattice parameter relaxation due to dopant incorporation, high resolution planar and cross-section transmission electron microscopy to view possible cluster signatures etc.

The lattice parameter enhancement was found to be similar to that realized in annealed film under cobalt incorporation. No cluster signatures could be noted from TEM studies. The details of this work will be published separately.[27]

The crystallinity and crystallographic orientation of the grown films were studied by X-ray diffraction (XRD) and transmission electron microcopy (TEM). From a XRD θ-2θ scan shown in Fig. 1, all the SRO, Co:TiO$_2$ and PZT films are (*00l*) oriented without any impurity phases. The in-plane epitaxy of all these layers was confirmed by four 90°-separated peaks in a XRD Φ scan measurement. TEM images in Fig. 2 show good crystallinity, no interdiffusion between the PZT and the Co:TiO$_2$ layers and no obvious signature of Co clustering in the Co:TiO$_2$ layer. A Rutherford backscattering (RBS) measurement further confirmed the composition for every layer and the clean interfaces between them. We note that magnetic force microscopy images for the films were mostly featureless with weak domain-like contrast possibly due to net weak magnetization. Occasional nanometer scale local dipole signatures could also be seen, which could arise due to surface topology defects or a local magnetic phase. Chambers et al.[28] have attributed the dipoles seen in their films to CoTiO$_3$ type ferromagnetic complexes. If we liberally calculate the concentration of such defects and attribute them entirely to such magnetic complexes[27], it still leads to ~ 4% cobalt dispersed uniformly in the matrix.

Ferroelectricity in the PZT layer was characterized using a RT-6000 system through the top and the bottom electrodes. Low frequency ferroelectric hysteresis measurements show a clear hysteresis loop as shown in the upper panel of the inset in Fig. 3 indicating the high quality of the PZT film. To measure the ferroelectric field effect on the Co:TiO$_2$ magnetization, an electric voltage pulse (1ms) was applied to the PZT layer through the top and bottom electrodes. The magnetization of the $Co:TiO_2$ layer was measured by superconducting quantum interference device (SQUID) magnetometry at room temperature after the electrical poling of all the top electrodes. During the magnetization measurement, the electrical voltage was turned off.

Fig. 3 shows the ferromagnetic hysteresis loops of the $Co:TiO_2$ layer measured after PZT poling. After poling of the PZT with a positive voltage, the ferromagnetic hysteresis loop of $Co:TiO_2$ was measured by SQUID, which showed a saturation magnetization value of about 100 μemu (squares in Fig.3). Then the PZT was negatively poled and the magnetization was measured again. The saturation magnetization value dropped to about 85 μemu (up-triangles in Fig. 3). After the PZT was poled by a plus voltage again, the saturation magnetization value went back to its initial value of 100 μemu (down-triangles in Fig. 3). After a second negative poling, the magnetization value dropped again to 87 μemu as shown by the circles in Fig.3. The saturation magnetization value as a function of applied voltage is plotted in the lower inset in Fig. 3. It is very clear that there are two stable states with a difference of about 15% in the magnetization of $Co:TiO_2$ and they can be switched by switching the polarization states of PZT.

The magnetization of the $Co:TiO_2$ film was also measured as a function of temperature as shown in Fig.4. The squares and up-triangles represent the data measured after a positive and a negative poling, respectively. In a large temperature range (150K – 400K), the difference between the two magnetization states persists at around 15% which is consistent with the hysteresis measurement at room temperature. The magnetization jump below 150K in the both curves is because of the magnetization of the SRO layer, which is ferromagnetic with a Curie temperature of 160K, and is not affected by the electric field.

A similar magnetization switching behavior was also measured on a PZT/Co:TiO$_2$ sample on Nb:SrTiO$_3$ substrate which is conducting and serves as the bottom electrode as well. The magnetization of Co:TiO$_2$ could once again be switched between two stable states indicating that the observed field induced magnetization modulation is a robust and reproducible result in different substrate-template layer configurations.

When the PZT is positively poled, its polarization points down inducing electron accumulation in the Co:TiO$_2$ layer, while negative poling depletes electrons in this layer. The ferroelectric field effect measurements on PZT/Co:TiO$_2$ samples on SRO-buffered-LAO and Nb:STO substrates clearly show that the magnetization is correspondingly increased (decreased) by increasing (decreasing) the charge carrier (electrons) density in Co:TiO$_2$. Thus there is a direct relationship between the magnetization and charge carriers in the Co:TiO$_2$ layer suggesting intrinsic nature of this DMS system. A rough estimate of the effective carrier density modulation can also be made as follows : Given the lattice parameters of anatase TiO$_2$ of a = b = 3.782 Å and c = 9.514 Å (2 molecules / cell), one has ~ 8.8 x 10$^{16}$ TiO$_2$ molecules in the 600 Å x 1cm x 1cm volume of the film. This yields ~ 6 x 10$^{15}$ Co atoms in this volume. Replacing Ti$^{4+}$ with Co$^{2+}$ will accommodate one oxygen vacancy and two carriers which are clearly localized as reflected by the insulating nature of the sample. Thus, assuming that all cobalt atoms are active as Co$^{2+}$ (which may be an over estimate) there would be a maximum of ~1.2 x 10$^{16}$ localized carriers in the film volume over 1 cm$^2$ area. The polarization charge obtained from the inset of Fig. 3 is ~ 60 μC/cm$^2$, which corresponds to surface density of ~ 3.5 x $10^{14}$ charges/cm$^2$. Since the sample is highly insulating, an equivalent of this charge will be induced or depleted as a function of polarity over the film volume across its thickness. This corresponds to about 3% minimum modulation of carrier density. Indeed if only a fraction (e.g. 4%) of cobalt is uniformly dispersed and active, and/or if some cobalt is in 3+ state, this estimate could even rise to about 6-8 %. As discussed below, within the framework of possible percolative mechanisms of ferromagnetism in diluted insulating system, this magnitude of modulation of carrier density is quite significant.

We now speculate on the possible origin of the observed field induced modulation of magnetization. Given the highly insulating nature of the samples no itinerant electron based picture is feasible, and therefore the RKKY-type scenario discussed extensively in the context of GaMnAs magnetization[18-19] simply does not apply here. Two other physical pictures which could conform naturally to the attendant insulating magnetic state may thus be relevant. One is the magnetic polaron percolation picture[29] discussed in the context of strongly insulating DMS materials, and the other is the defect (F-center) state percolation model discussed in the context of the large ordered moment in Fe doped SnO$_2$ system[30]. We believe that our insulating Co:TiO$_2$ system manifesting the electric field modulated ferromagnetism could actually be a combination of these two percolation scenarios.

In the polaron percolation picture[29], each local magnetic moment (the local moment bound to carriers) is the effective magnetic polaron and the whole system is a collection of a random distribution of these polarons. At T$_C$, the bound magnetic polarons form a percolating path leading to global ferromagnetism. (The magnetic percolation does not imply transport percolation since each polaron is strongly localized). Subjecting such a system to strong external electric field could lead to electric dipolar distortion of the magnetic polaron with attendant distortion of the shape of the corresponding wave function and field driven redistribution effects. The corresponding changes in the geometry of the percolation network should change the magnetic moment of such a system.

In the model suggested by Coey et al. [30], the reason for ferromagnetism in some of these dilutely doped compounds is envisioned to be due the F-center exchange (FCE). As a lower valent cation ($Co^{2+}/Co^{3+}$) replaces a higher valent site (such as $Ti^{4+}$ or $Sn^{4+}$), oxygen vacancies are expected to be created to account for the local charge balance. Three types of oxygen vacancies can occur : a) vacancy with no electrons present (~ net +2 charge), b) vacancy with one electron trapped (~ net +1 charge) – termed as $F^+$ center and c) vacancy with two electrons trapped and no net charge (F-center). The diameter of the F-center carrier wave function depends on the dielectric constant of the host and this is between 1-2 nanometers for high dielectric constant materials such as $TiO_2$ or $SnO_2$. Overlap of the d-shell orbitals of the magnetic dopant with the perturbation-orbital can then lead to a *co-operative percolative ordering* of such magnetic ions even in the absence of (mediating) itinerant carriers. When this physical picture is visualized under application of strong electric field, once again the perturbation-orbital should undergo distortion changing the microscopic constitution of the percolation and therefore the magnetic moment. In an even higher electric field, ionization of some F-centers can be envisaged which would clip magnetic linkages and weaken the magnetism. In fact one could envision each of this F-center induced magnetic moment as an effective bound polaron, which then forms a percolating cluster leading to global ferromagnetism along the line of the magnetic polaron percolation picture of Ref. 29. An external electric field will affect global ferromagnetism by strongly distorting each polaron.

In conclusion, we have demonstrated for the first time a reversible external electric field induced modulation of room temperature magnetic moment in an oxide-based diluted magnetic semiconductor. A strong modulation of ~15% is realized in a high quality epitaxial heterostructure of $PbZr_{0.2}Ti_{0.8}O_3/Co:TiO_2/SrRuO_3$ grown on (001) $LaAlO_3$. Observation of electric field effect strongly favors intrinsic (i.e. carrier-induced) nature of ferromagnetism in the high temperature grown insulating $Co:TiO_2$ films. Our finding is also the first report of field effect on ferromagnetism in an *insulating* diluted magnetic system. An ability to control ferromagnetism by an electric field is promising for the next generation electronic devices for information storage and memory applications.

This work was supported by DARPA SpinS program (through US-ONR) and the NSF-MRSEC (DMR 00-80008) at Maryland. The PLD and RBS facilities used in this work are shared experimental facilities (SEF) supported in part under NSF-MRSEC. The authors would like to thank R. L. Greene for critical reading of the manuscript.

Figure Captions

Fig.1 X-Ray Diffraction θ-2θ scan of the deposited PZT/Co:$TiO_2$/SRO/LAO multilayers. The inset is the schematic structure of the sample.

Fig.2 Transmission electron microscopy images of the PZT/Co:$TiO_2$/SRO/LAO sample.

Fig.3 Magnetic hysteresis loops of the Co:$TiO_2$ layer after several electric voltage poling on the PZT layer. The upper panel of the inset shows the ferroelectric hysteresis loops of the PZT layer, while the lower panel is the saturation magnetization of Co:$TiO_2$ as a function of the applied voltage on PZT.

Fig.4 Temperature dependence of magnetization of Co:$TiO_2$ layer measured under a magnetic field of 1300 Oe after positive and negative poling on PZT.

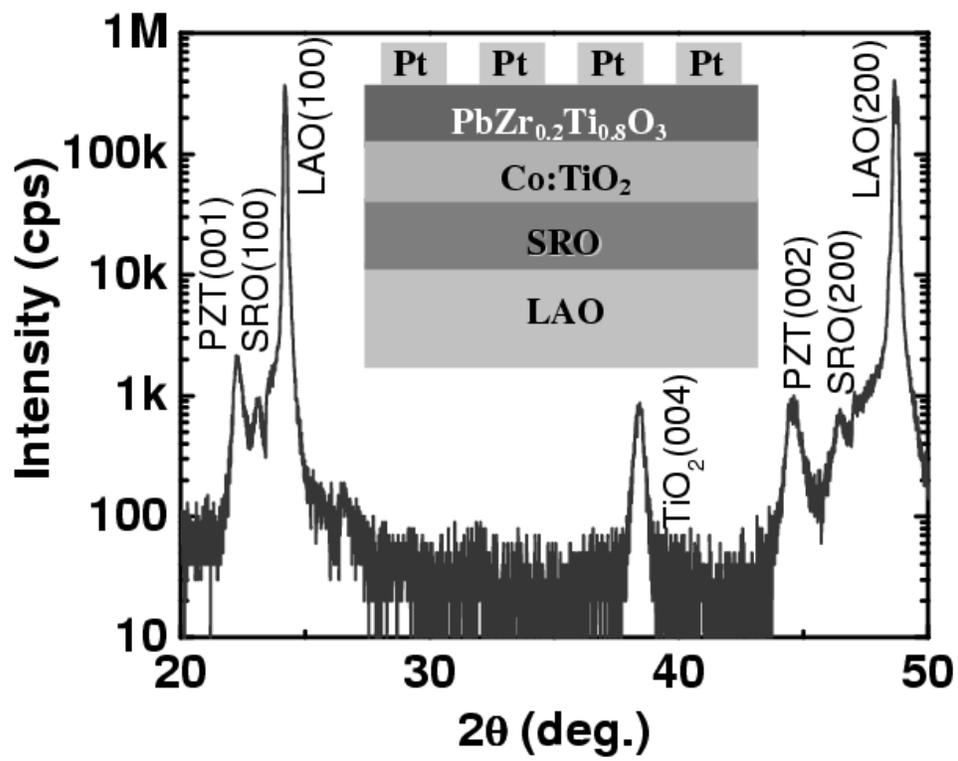

Figure 1

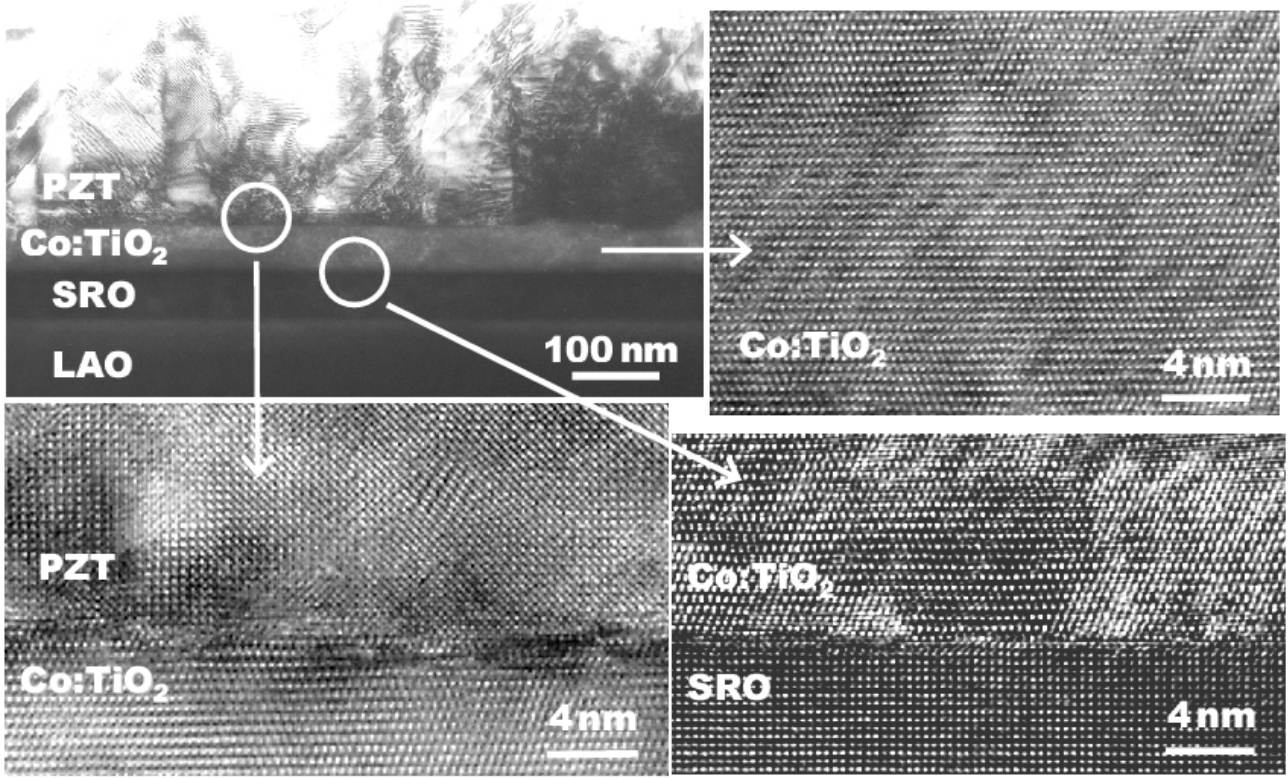

Figure 2

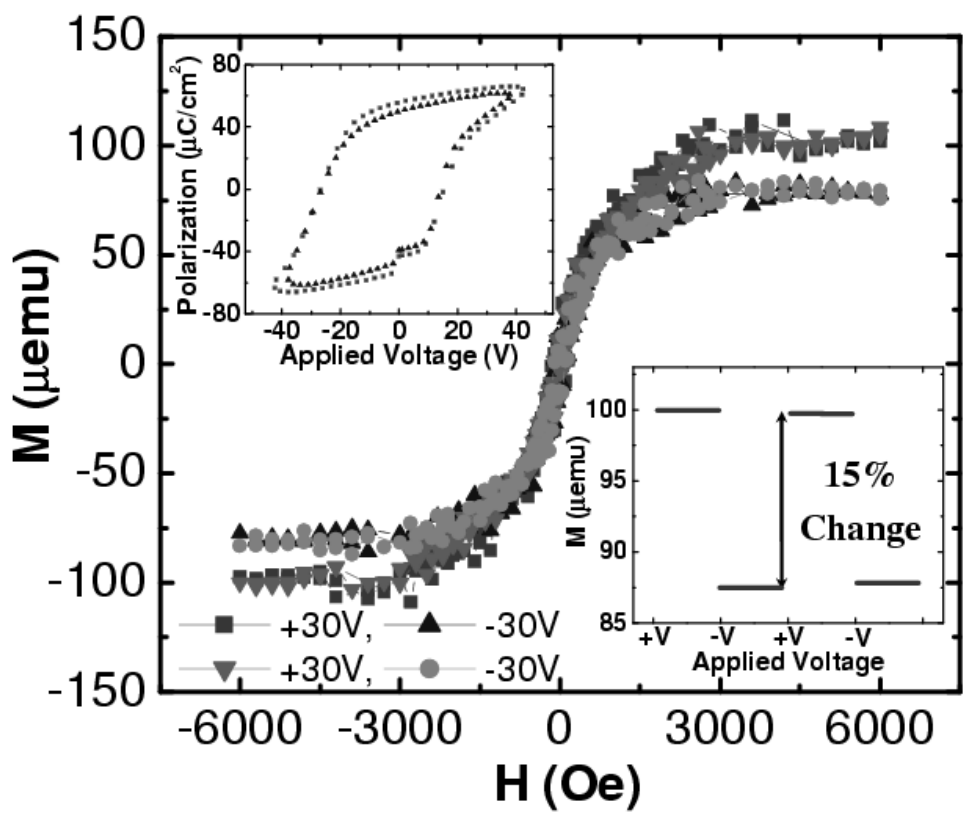

Figure 3

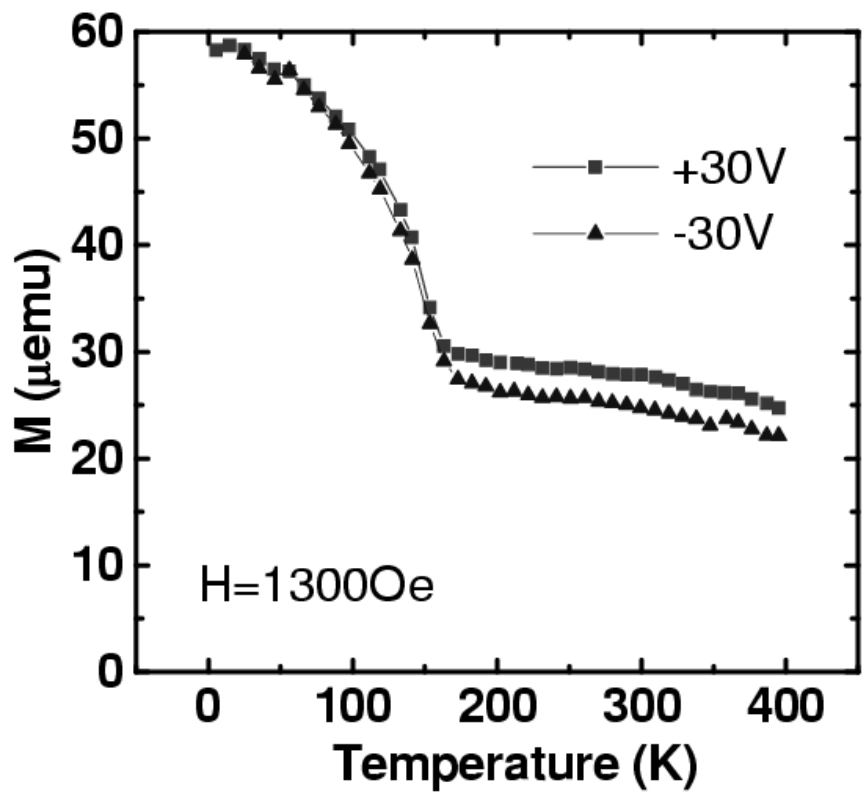

Figure 4